\documentclass[lettersize,journal]{IEEEtran}
\usepackage{amsmath,amsfonts}
\usepackage{algorithmic}
\usepackage{cite}
\usepackage{array}
\usepackage[caption=false,font=normalsize,labelfont=sf,textfont=sf]{subfig}
\usepackage{textcomp}
\usepackage{stfloats}
\usepackage{url}
\usepackage{verbatim}
\usepackage{graphicx}
\usepackage{multirow}
\usepackage{hyperref}
\usepackage[table,xcdraw]{xcolor}
\def\BibTeX{{\rm B\kern-.05em{\sc i\kern-.025em b}\kern-.08em
		T\kern-.1667em\lower.7ex\hbox{E}\kern-.125emX}}
\usepackage{balance}
\begin{document}
	\title{\huge {Rotatable Antenna Enabled Wireless Communication and Sensing: Opportunities and Challenges}}
\author{ 
	Beixiong Zheng,~\IEEEmembership{Senior Member,~IEEE},
	Tiantian Ma,  
	Changsheng You,~\IEEEmembership{Member,~IEEE}, \\
	Jie Tang,~\IEEEmembership{Senior Member,~IEEE},
	Robert Schober,~\IEEEmembership{Fellow,~IEEE}, 
	and Rui Zhang,~\IEEEmembership{Fellow,~IEEE}

\thanks{
	Beixiong Zheng, Tiantian Ma, and Jie Tang are with the South China University of Technology; 
	Changsheng You is with the Southern University of Science and Technology;
	Robert Schober is with the Friedrich-Alexander University of Erlangen-Nuremberg;
	Rui Zhang (corresponding author) is with the National University of Singapore.
}
}
	
	
	\maketitle 
	
\begin{abstract}
Rotatable antenna (RA) is an emerging technology that offers significant potential to enhance wireless communication and sensing performance by flexibly adjusting the boresight of directional antennas. Specifically, RA can flexibly reconfigure its boresight direction via mechanical or electronic means, thereby improving communication channel conditions and/or enhancing sensing resolution and range. 
In this article, we first provide an overview of RA, covering its hardware architectures and radiation pattern characterization.
We then illustrate how RA improves communication performance through interference mitigation, spatial multiplexing, and flexible beamforming, as well as sensing capabilities in terms of coverage, resolution, and multi-target/dimensional sensing. Furthermore, we highlight representative applications of RA and discuss key design challenges in RA systems, including rotational scanning scheduling, channel estimation/sensing, boresight optimization, and RA configuration.
Finally, both experimental and simulation results are provided to validate the performance gains achieved by RA for both communication and sensing. Leveraging its unique capabilities in flexible antenna/array rotation to adapt to various communication/sensing requirements and channel conditions, RA is poised to become a key enabler of future intelligent, resilient, and agile wireless networks.
\end{abstract}
	
%

\section{Introduction}
    As global efforts devoted to developing the sixth-generation (6G) wireless networks intensify, future wireless systems are expected to enable ubiquitous sensing and massive connectivity with unprecedented accuracy and capacity. This vision imposes stringent demands for higher spectral efficiency and spatial resolution far beyond that of current multiple-input multiple-output (MIMO) systems. On one hand, while existing MIMO systems attempt to enhance their performance by scaling up the number of antennas, such array enlargement inevitably incurs prohibitively high hardware costs and deployment/processing complexity. This issue also poses challenges for emerging applications which increasingly call for device miniaturization and operational agility, such as Internet-of-Things (IoT) and unmanned aerial vehicles (UAVs). On the other hand, current MIMO systems mostly employ fixed-antenna arrays, which lack the spatial flexibility to dynamically adapt to channel variations, leaving much of the channel spatial diversity underutilized. As such, how to boost MIMO system capacity and spatial resolution within a constrained array aperture by fully harnessing the spatial degrees-of-freedom (DoFs) becomes the focus of future research \cite{ying2024reconfigurable}.

    Different from conventional antenna number scaling paradigms, (non-fixed) flexible antenna architectures have emerged as promising alternatives for enhancing wireless system capacity by dynamically reconfiguring antenna characteristics, such as position, shape, and orientation, in response to spatial channel variations. 
    Among these architectures, fluid antenna system (FAS) \cite{wong2021fluid} and movable antenna (MA) \cite{zhu2024modeling} have attracted considerable interest because of their ability to adjust antenna shape and/or position to create more favorable channel conditions.
    However, their performance in practice is often constrained by limited movement speed and considerable hardware response latency. To unlock additional spatial DoFs, six-dimensional MA (6DMA) extends antenna reconfigurability to both three-dimensional (3D) position and rotation adjustments \cite{shao20256d}. While this additional flexibility significantly enhances the adaptability to the wireless environment, it requires increased design complexity and implementation cost due to the intricate geometric and mechanical constraints for antenna movement.

    Motivated by the above, rotatable antenna (RA) \cite{zheng2025rotatable} has emerged as a viable solution to enhance spatial DoFs while maintaining a compact aperture and low hardware cost.\footnote{Historically, RA is not a new concept. Since mechanically controlled RAs originated in the mid-20th century radar systems, RAs have undergone a transformative evolution. Over time, the use of electronic controls and more efficient motors has significantly enhanced the flexibility and performance of RAs. Earlier research primarily focused on the orientation adjustment of entire antenna arrays, with applications largely in radio navigation and localization\cite{jiang2010localization}. In contrast, the RA considered in this article enables independent boresight direction adjustment of individual antennas within an array, thus offering greater flexibility, more intelligence, and higher performance gains for next-generation wireless systems.} 
    By independently adjusting the 3D boresight direction of each antenna via mechanical or electronic means, RA can rotate the radiation pattern towards desired directions, thereby achieving significant performance gains in terms of precise beam alignment, extended coverage, and adaptive interference management.
    Table~\ref{tab:sum} summarizes and compares the key characteristics of representative flexible antenna architectures, including FAS, MA, 6DMA, flexible antenna array (FAA) \cite{yang2025flexible}, and RA.
    Although RA \cite{zheng2025rotatable} can be considered as a special case of 6DMA\cite{shao20256d, shao20256dmovable} in terms of movement model by sharing the common rotational flexibility, their practical architectures and implementations differ significantly.
    In particular, unlike the translational movement in MA/6DMA, which typically needs additional space (e.g., sliding tracks or movable area), RA only requires local rotational adjustment, which can be achieved more easily through compact mechanical or integrated electronic design. 
    This makes RA a streamlined and scalable flexible-antenna solution, offering great compatibility with existing wireless systems as well as enabling efficient deployment in space and cost-constrained scenarios such as IoT nodes and wearable devices.

\begin{table*}[]
	\centering	
	\caption{Comparison of Different Flexible Antenna Architectures}
	\vspace{-0.2cm}
	\label{tab:sum}
	\renewcommand{\arraystretch}{1.5}
	\begin{tabular}{|c|c|c|c|c|}
		\hline
		\textbf{\begin{tabular}[c]{@{}c@{}}Antenna \\ Architecture\end{tabular}} &
		\textbf{Hardware  Implementation} &
		\textbf{\begin{tabular}[c]{@{}c@{}}Reconfigurable Parameter\end{tabular}} &
		\textbf{\begin{tabular}[c]{@{}c@{}}Deployment \\ Complexity\end{tabular}} &
		\textbf{\begin{tabular}[c]{@{}c@{}}System\\ Overhead\end{tabular}} \\ \hline\hline
		RA \cite{zheng2025rotatable} &
		\begin{tabular}[c]{@{}c@{}}Motors or electronic means\end{tabular} &
		\begin{tabular}[c]{@{}c@{}}Antenna 3D boresight direction\end{tabular} &
		Very Low &
		Very Low \\ \hline
		FAA \cite{yang2025flexible} &
		Flexible materials &
		Array shape/structure &
		Moderate &
		Moderate \\ \hline
		FAS \cite{wong2021fluid} &
		Pixel antennas or liquid materials &
		\begin{tabular}[c]{@{}c@{}}Antenna shape and 3D position\end{tabular} &
		Moderate &
		Moderate \\ \hline
		MA \cite{zhu2024modeling} &
		Motors &
		Antenna 3D position &
		Moderate &
		Moderate \\ \hline
		6DMA \cite{shao20256d} &
		Motors and flexible cables &
		\begin{tabular}[c]{@{}c@{}}Antenna 3D position and 3D rotation\end{tabular} &
		Moderate &
		Moderate \\ \hline
	\end{tabular}
\end{table*}

\begin{figure*}[!t]
	\centering
	\includegraphics[width=0.8\textwidth]{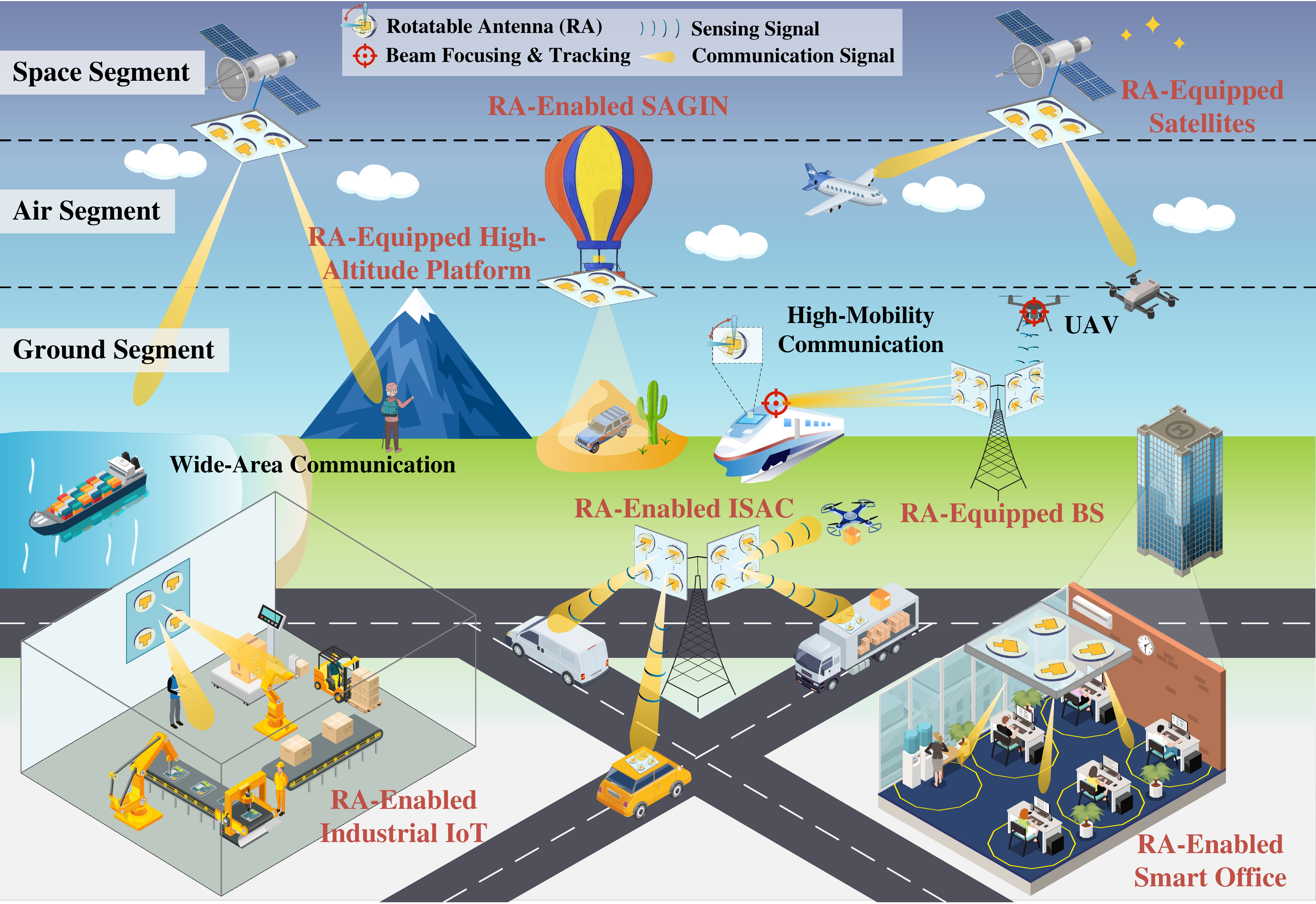}
	\caption{Typical scenarios for RA-enabled wireless communication and sensing.}
	\label{scenes}
\end{figure*}

    Given its unique ability to dynamically adjust the antenna boresight in 3D space, RA technology enables a fundamental paradigm shift in base station (BS) coverage from purely ground-facing service  to seamlessly supporting both terrestrial and aerial users/targets. 
    As illustrated in Fig.~\ref{scenes}, RA opens new opportunities for a broad range of future wireless applications, which will be further elaborated in Section~\ref{sec:application}. Motivated by the above, this article aims to provide a comprehensive overview of RA-enabled wireless communication and sensing, with an emphasis on the fundamental principles, hardware architectures, key performance advantages, potential applications, and practical challenges. Through both experimental and simulation results, we also demonstrate how RA technology can significantly enhance communication and sensing performance of wireless systems under practical constraints.

    \section{Hardware Architecture and Radiation Pattern Characterization}
\begin{figure*}[!t]
	\centering
	\includegraphics[width=1\textwidth]{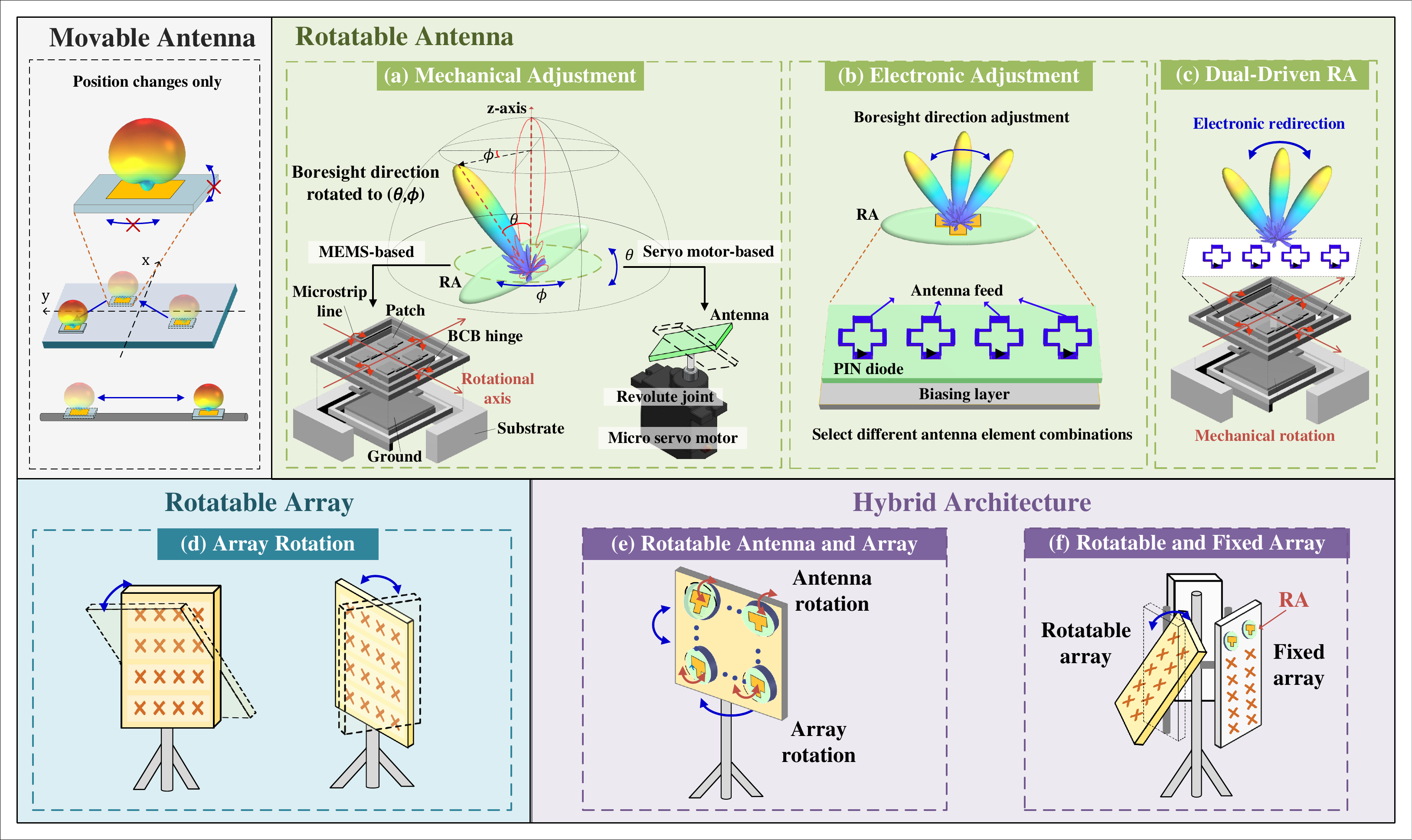}
	\caption{Architectures and hardware implementations for RA.}
	\label{hardware}
\end{figure*}

    Fig.~\ref{hardware} illustrates various hardware architectures for implementing RAs. In addition to the communication module (including radio frequency chain, baseband processing unit, etc.) present in conventional fixed-antenna architectures, RA is incorporated with a dedicated control module for rotating the boresight direction of each antenna or antenna array. Specifically, this boresight adjustment can be achieved through either mechanical or electronic mechanisms, as detailed below.
    
\begin{itemize}
    \item As illustrated in Fig.~\ref{hardware}(a), mechanically driven RA employs actuators to physically rotate the antenna orientation, thereby altering its boresight direction. A typical implementation involves mounting the RA on a servo-motor-controlled platform, where the platform's rotation directly adjusts the antenna's orientation in 3D space \cite{dai2025a,dai2025rotatable}. Alternatively, mechanical adjustment can be achieved using micro-electromechanical system (MEMS) technology to facilitate miniaturization and simplify deployment \cite{baek2017a}. In such systems, torsion hinges are commonly employed to enable high-precision and robust rotational control of the antenna, offering advantages including compact size, rapid actuation, and low power consumption.
	\item As illustrated in Fig.~\ref{hardware}(b), electronically driven RA maintains fixed antenna orientation, while its boresight direction is dynamically reconfigured through electronic techniques. For example, for a multi-feed antenna, its main lobe can be rotated in different directions by activating different feed points on a directional radiator \cite{morshed2023beam}. An alternative approach involves the use of parasitic elements with electronic tuning. By tuning the loads (e.g., varactor diodes, PIN diodes) of these parasitic elements, the direction of the antenna radiation pattern can be rotated accordingly \cite{mechael2021electronically}. However, these techniques mainly offer discrete and predefined radiation directions. To further enable continuous boresight direction adjustment, more advanced tunable materials (e.g., liquid crystal \cite{lindle2016efficient}) can be used.
\end{itemize}
    
    In general, mechanically driven methods offer a broader control range for boresight adjustment, while electronically driven methods provide better compatibility with existing wireless platforms.
    In practice, MEMS-based actuators typically entail a power consumption in the milliwatt range, with response time on the order of microseconds to milliseconds. Meanwhile, electronically driven RAs can typically achieve response time ranging from nanoseconds to milliseconds.
    To harness the benefits of both approaches, co-designed RA architectures that integrate both control mechanisms can be adopted to enable wide-angle and rapid reconfiguration of the radiation pattern, as depicted in Fig.~\ref{hardware}(c).
 
    In addition to antenna-level rotation, mechanically driven RA systems can also incorporate array-level rotation, as illustrated in Fig.~\ref{hardware}(d).
    Independently rotating each antenna in an array allows for finer-grained and more flexible boresight reconfiguration. However, this comes at the cost of increased control overhead and design complexity compared to rotating the entire array as a unit. On the other hand, array-level rotation inherently alters the antenna radiation pattern, which may require additional compensation mechanisms such as antenna rotation calibration to maintain the desired system performance. Consequently, dual-scale architectures with both antenna-level and array-level rotation can be adopted to further enhance rotation resolution and range, as shown in Fig.~\ref{hardware}(e). Furthermore, to enhance compatibility with existing BS architectures, hybrid configurations that integrate RAs into fixed-antenna BSs can be adopted, as depicted in Fig. ~\ref{hardware}(f). In this setup, the sector antennas retain their inherent capability to serve terrestrial users, while the co-deployed antenna/array-level RAs extend coverage to serve both terrestrial and aerial users/targets. This design enhances the system's overall connectivity and coverage performance, particularly benefiting emerging low-altitude economy with rich UAV applications.
 
    Besides hardware architecture, a fundamental aspect of RA is the characterization of how its radiation pattern is manipulated during operation. As shown in Fig.~\ref{hardware}, both MA and RA preserve the intrinsic shape of their antenna radiation patterns during reconfiguration, but they manipulate the pattern through fundamentally different mechanisms. Specifically, MA translates the radiation pattern by relocating the antenna's position, whereas RA rotates its radiation pattern by adjusting the antenna's or array's boresight direction, thereby enabling angular diversity to enhance communication and sensing performance. Nevertheless, the radiation pattern rotation flexibility in practice depends on implementation feasibility and environmental constraints. For example, high rotation resolution demands either a high-precision actuator or a large number of feasible switching or excitation states, both of which incur substantial hardware cost. Moreover, the rotational range of RA also depends on the adjustable range of the control system and the surrounding environment. By selecting an appropriate control accuracy and range, the radiation pattern of RA can be effectively rotated to allocate signal power towards desired directions, thereby enhancing directional gain and suppressing interference.

\section{Performance Advantages of RA}
    In this section, we highlight the performance advantages of RA over fixed-antenna architectures for wireless communication and sensing in terms of interference mitigation, spatial multiplexing, flexible beamforming, as well as sensing resolution/coverage, and multi-target/dimensional sensing.
    
\subsection{Wireless Communication}

\begin{itemize}
	\item \textbf{Interference mitigation:} Conventional fixed-antenna systems primarily rely on resource allocation and signal processing techniques such as filtering, equalization, precoding/combining to suppress or cancel interference. While effective, these techniques impose significant signal processing overhead and may suffer performance degradation under strong interference from dominant sources. In contrast, RA can exploit its boresight direction reconfigurability to mitigate interference in a more direct and efficient manner. Since interference signals often arrive from different directions, RA can rotate its main lobe to suppress interference while preserving high gain towards intended users. This capability is particularly beneficial for cellular networks, where co-channel interference and multi-path fading are inevitable. By adaptively rotating its radiation pattern in 3D space, RA enables efficient interference mitigation to enhance link quality and reliability with less signal processing overhead.
	\item \textbf{Spatial multiplexing:} In rich scattering environments, MIMO systems enhance channel capacity by spatially multiplexing multiple data streams for simultaneous transmission. The achievable spatial multiplexing gain hinges on rich scattering which enables receive antennas to effectively distinguish signals from different transmit antennas. However, this channel condition may not be achieved in practice when there exists  a dominant line-of-sight (LoS) component or strong reflected paths, especially in near-field scenarios, resulting in a reduced effective channel rank and, consequently, degraded achievable multiplexing gain. To alleviate such performance degradation, RA provides a new means to create subchannel orthogonality and thereby improve the channel conditions. 
	Through dynamic boresight adjustment, RA arrays enhance the angular separation among signal paths, which improves spatial resolvability and facilitates concurrent transmissions of multiple data streams in both point-to-point multi-user communication systems.
	\item \textbf{Flexible beamforming:} Besides the capacity gain achieved by spatial multiplexing, RA arrays can also be utilized to enable more flexible beamforming. Traditionally, beamforming is realized by adjusting the complex weights across antennas to steer the entire array pattern towards desired directions, which we refer to as array-level beamforming. In contrast, RA arrays can reconfigure the radiation pattern of each individual antenna to achieve element-level beamforming for more precise and flexible far-field beam steering or near-field beam focusing towards a specific direction/location. As a result, by jointly optimizing the array beamforming weights and the antenna boresight directions, RA arrays offer higher beamforming gains by enabling more flexible power distribution over beam directions compared to conventional fixed-antenna arrays.
\end{itemize}

\subsection{Wireless Sensing}
In addition to enhancing communication performance, RA's unique beam tracking/scanning capability provides additional benefits for wireless sensing applications.

\begin{itemize}
	\item \textbf{Sensing resolution:} Wireless sensing resolution is closely related to the beamwidth, antenna aperture, and signal processing techniques. Conventional fixed-antenna systems rely on array processing to sharpen the angular response of the array to improve the angular discrimination between closely spaced targets. In contrast, the RA array introduces an additional DoF by enabling the mechanical or electronic alignment of the boresight direction of each antenna. When all RAs coherently direct their boresights towards a common spatial point (e.g., the target position), the array effectively acts as a spotlight for near-field sensing, thereby concentrating energy not merely in a narrow angular sector but at a specific point in 3D space. Therefore, by effectively combining near-field beam focusing and boresight alignment, RA arrays offer high directional gain and enhanced spatial resolution for wireless sensing applications.
	\item \textbf{Sensing coverage:} Besides spatial resolution enhancement, the ability of RA to flexibly adjust its boresight direction significantly extends the spatial coverage of wireless sensing systems. Traditional fixed antennas rely on predetermined radiation patterns, which often result in limited spatial coverage and blind zones. In contrast, RA can adjust its boresight direction over time across a much wider 3D space to actively scan the region of interest, enabling more reliable detection of mobile targets and wide-angle coverage. Moreover, in hybrid configurations where RAs are co-deployed with fixed-antenna arrays, the flexible redirection of RAs can be leveraged to complement fixed-antenna coverage, thereby effectively mitigating blind spots and supporting robust sensing in distributed or irregular target scenarios.
	\item \textbf{Multi-target/dimensional sensing:} In many wireless sensing applications, it is necessary to detect multiple targets or extract multi-dimensional information simultaneously. However, this task  becomes highly challenging with a limited number of antennas, as closely spaced targets often interfere with each other, leading to signal space overlap and reduced spatial resolvability. RAs provide an effective solution to address this issue by allowing independent boresight adjustment of each antenna. This enables the RA array to sequentially or concurrently scan multiple spatial directions using fewer antennas, enhancing its ability to resolve and distinguish signals from closely spaced targets. Similarly, through diverse angular observations via boresight adjustment, RAs can extract multiple target features (e.g., direction, velocity, size, orientation), thereby facilitating cost-effective multi-dimensional sensing.
\end{itemize}

\section{Potential Applications of RA}
\label{sec:application}
Building on the  aforementioned performance advantages, RA enables a wide range of promising applications in future wireless systems. In this section, we highlight several representative scenarios where the boresight reconfigurability of RA can be effectively leveraged to meet increasingly stringent requirements in terms of capacity, coverage, spatial resolution, and other key system performance metrics for emerging applications.
\begin{itemize}
	\item \textbf{Machine-type communication (MTC):} Given the demand for massive device connectivity and ultra-reliable low-latency communication, MTC has become a fundamental pillar of future wireless networks, particularly for emerging applications such as smart cities and industrial automation.
	However, meeting this demand in practice remains challenging, especially considering the dense node deployment and small-payload devices. In this context, the RA system can adjust antenna's boresight direction based on the spatial distribution of MTC devices, enabling more effective utilization of the spatial DoFs to enhance signal quality and suppress interference. Furthermore, RA provides a lightweight and scalable solution for MTC devices to deliver high-performance communication and sensing without necessitating large-scale antenna arrays. 
	\item \textbf{Space-air-ground integrated network (SAGIN):} By seamlessly integrating the space, air, and ground segments, SAGIN aims to deliver global coverage and low-latency communication for future 6G systems.
	However, conventional down-tilted BSs struggle to support aerial and spaceborne nodes, and their fixed antenna/array orientations further impede the maintenance of reliable links for the dynamic topologies created by mobile aerial and space platforms. To overcome these challenges, RA arrays can be co-deployed at existing BSs to extend communication/sensing coverage across a wider 3D space, thereby enhancing connectivity and tracking capability of the airborne and space segments. Alternatively, RA arrays can be deployed on the aerial or spaceborne nodes to enable flexible beam control, allowing them to adaptively track and maintain high-quality links with ground nodes.
	\item \textbf{Indoor sensing and localization: } Indoor sensing and localization are critical for emerging intelligent applications such as smart homes and robotic navigation. Unlike outdoor environments, indoor scenarios usually suffer from severe LoS blockage caused by walls, furniture, and other obstacles, which significantly degrades sensing and localization accuracy. This limitation, however, can be alleviated by employing RAs, which rotate their radiation patterns to enhance the directional gain towards LoS reflected paths. Beyond localization, the reconfigurability of RAs is advantageous for simultaneous localization and mapping (SLAM) in robotic networks.
	By adaptively adjusting the boresight directions, the RA system can efficiently construct channel knowledge maps to facilitate environment-aware communications and sensing with improved accuracy and robustness.
	\item \textbf{Integrated sensing and communication (ISAC):} ISAC serves as a key enabler of 6G networks by integrating high-rate communications and high-precision sensing into shared infrastructure and spectrum.
	However, a key challenge arises when the spatial directions of communication users and sensing targets differ, leading to conflicts in beam steering and resource allocation. To address this issue, RA offers a promising solution by enabling independent 3D boresight control across different antennas in the array. Specifically, RAs can be directed towards different sensing regions to enhance angular resolution and/or aligned with different communication users to improve link quality. Furthermore, by adaptively allocating antenna resources based on the spatial distribution of users and targets, RA can also enhance coverage efficiency in cluttered or dynamic environments.
\end{itemize}
	
\section{Design Challenges: Unlocking the Full Potential of RA}
Despite the appealing advantages and significant potential of RA, new design challenges need to be addressed for RA-enabled wireless communication and sensing. In the following, we elaborate on the main challenges in rotational scanning scheduling, channel estimation/sensing, boresight optimization, and antenna configuration. In addition, we provide forward-looking solutions for future exploration.

\subsection{Rotational Scanning Scheduling}
Scanning scheduling plays a pivotal role in RA systems, as it directly governs the trade-off between key communication/sensing performance metrics and overall system overhead. By virtue of its reorientation capability, RA can scan across multiple directions to acquire environmental information over a wide angular range. This capability helps identify and track targets with high mobility, thereby facilitating wide-area sensing/estimation and adaptive boresight optimization. Despite these advantages, the rotational scanning process inevitably introduces control overhead and latency, which may degrade sensing accuracy and reduce data rates under dynamic channel or mobility conditions. This highlights the need for carefully designed rotational scanning schedules.

Generally, rotational scanning scheduling poses several key challenges. First, the design of scanning scheduling must account for multiple factors, including the scanning parameters (e.g., range, frequency, pattern) as well as the coordination with other signal processing tasks to ensure efficient system operation. This makes the scheduling problem inherently intricate. Second, different tasks, such as communication, sensing, navigation, and localization, impose diverse requirements on scanning resolution, latency, and angular precision, thereby necessitating adaptive and context-aware scheduling strategies.
To address these challenges, 
several effective solutions can be adopted. In cases where prior information is available, such as channel statistics or environmental information, RA can adaptively tailor its scanning pattern and range to reduce redundant scans. On the other hand, when such prior information is unavailable or unreliable, a coarse-to-fine scanning strategy can be adopted to effectively balance scanning resolution and latency. To further reduce the scanning overhead, multi-RA cooperative scheduling designs can be employed. In multi-RA systems, dividing the antennas into groups for performing different scanning tasks can help reduce the workload per RA while maintaining full coverage, which is particularly useful in large-scale deployments or distributed sensing networks. 
	
\subsection{Channel Estimation/Sensing}
Accurate channel state information (CSI) is indispensable for enabling effective boresight control and unlocking the full potential of RA systems. Unlike FAS and MA/6DMA systems, which typically require reconstructing the full CSI associated with the entire antenna movement region for position optimization, RA systems only need to estimate a small number of environmental parameters for channel estimation or sensing similar to fixed-antenna MIMO. This advantage stems from the fact that RA simply rotates the radiation pattern while maintaining fixed antenna positions during boresight adjustment, thereby preserving the physical geometry of the transceiver setup and propagation environment. In this regard, conventional channel estimation techniques remain applicable in RA systems, ensuring their compatibility with existing wireless infrastructures.

Beyond this baseline compatibility, RA systems offer new opportunities as well as challenges for coverage/accuracy enhancement in estimation and sensing by actively exploiting the spatial DoFs. By sequentially or adaptively scanning the dynamic wireless environment, RAs can achieve multi-view estimation or sensing, thereby enabling more accurate multi-path resolution and/or target separation. However, while these multi-view observations obtained over time can effectively enhance estimation or sensing resolution, they also require more complex space-time signal processing techniques.
To address this, it is essential to develop low-overhead estimation schemes that leverage prior geometric knowledge and spatio-temporal correlations in the environment.
A promising approach in this context is to adopt learning-based frameworks, such as neural networks, to capture correlations across boresight directions, thereby enhancing the estimation and sensing performance for high-dimensional observations or under noisy measurement conditions.

\subsection{Boresight Direction Optimization}
For RA systems, finding the optimal boresight direction of each antenna is critical to enhancing communication and sensing performance. For instance, in the single-user communication or single-target sensing scenario, the boresight directions of all RAs and the array beamforming vector can be jointly designed to maximize the received signal power in the user/target direction. This enables the RA system to achieve not only a beamforming gain as in conventional fixed-antenna systems but also a near-field beam-focusing gain via boresight alignment. For the more general multi-user/multi-target setup, RA systems can benefit not only from the enhanced directional gain or sensing resolution for the desired directions but also from the suppression of co-channel interference.

However, boresight direction optimization also presents several unique challenges. 
One challenge in RA boresight direction design lies in the practically available discrete boresight direction levels of each RA. Instead of exhaustively searching over all possible discrete levels of RAs, a practical approach is to first relax discrete constraints and solve the problem for continuous boresight direction values, and then quantize the obtained solutions to their nearest values in the discrete set. 
Alternatively, heuristic algorithms like greedy search can be employed to efficiently explore the discrete solution space with low complexity.
On the other hand, the effectiveness of boresight direction optimization heavily relies on the availability of accurate CSI, which is challenging to acquire in practice. To address this issue, CSI-robust alternatives such as hierarchical search can be adopted, where each antenna first performs coarse scanning to identify promising directional regions and then refines its search within such regions. As for the case without explicit CSI, machine learning-based methods, such as reinforcement learning and neural networks, can be exploited. 

\subsection{Antenna Configuration}
From concept to implementation, the configuration of RA plays an essential role in determining both the system performance and hardware efficiency. 
While RA offers a compact and scalable solution to enhance spatial flexibility, configuring it for practical use remains challenging.
In particular, key design aspects, such as boresight control methods, rotational scale, array structure, and deployment strategy, directly affect control accuracy, latency, size, and cost, thereby impacting the overall performance and integration complexity of RA systems.
These considerations become even more critical when RAs are deployed on spaceborne or aerial platforms, where additional requirements for miniaturization and ruggedization need to be addressed.

Generally, the optimal RA configuration largely depends on the system tasks. In wireless sensing systems that prioritize spatial resolution, sparse RA arrays are often desirable, as their large array aperture enables finer sensing resolution. Moreover, the increased inter-antenna spacing helps mitigate coupling effects during the boresight rotation. In communication systems, however, a sparse array structure may cause severe inter-user interference when users fall into each other's beam grating lobes. This underscores the importance of context-aware evaluation when applying sparse RA arrays in ISAC systems. 
Beyond system functionality, environmental factors further influence RA configuration strategies.
For instance, in rich-scattering or sparsely populated environments, distributed RA deployment is preferred to enhance coverage performance and provide macro diversity gains to reduce detrimental shadowing effects.
In aerial or spaceborne platforms that are typically size, weight, and power-constrained, compact arrays with small antenna size and electronically driven or MEMS-actuated rotation methods are preferred for practical integration.
Furthermore, such platforms are often subject to mechanical vibrations, jitters, and thermal variations, which can induce pointing errors and consequently degrade system performance \cite{hayal2023modeling}. To mitigate these effects, ruggedization and robust hardware design strategies should be incorporated.

\begin{figure}[!t]
	\centering
	\includegraphics[width=3in]{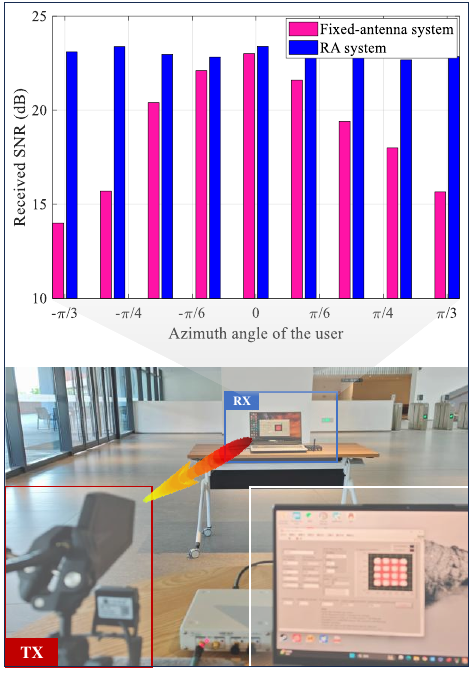}
	\caption{The received SNR versus the azimuth angle of the user. (Demonstration videos are available
		at {\href{https://youtu.be/r4zzDOmQ2ZI}{https://youtu.be/r4zzDOmQ2ZI}} and {\href{https://youtu.be/L1aMVaGj5rw}{https://youtu.be/L1aMVaGj5rw}}.)}
	\label{single_comm}
\end{figure}

\section{Case Studies}

To validate the performance advantages of RA in wireless networks, we present experimental and simulation results for RA-enabled communication/sensing. We first conduct a practical experiment to evaluate the coverage performance in the single-user communication case using a mechanically driven RA prototype. Subsequently, simulations are performed to validate the multi-user communication and multi-target sensing performance gains achieved by RA.

\subsection{Experimental Results: Single-User Communication}
The experiment is conducted in an indoor environment, where an RA-equipped transmitter operates at a carrier frequency of $5.8$ GHz using 16-quadrature amplitude modulation (QAM). The transmit power is set to $5.8$ dBm, and the data rate is $0.5$ Mbps. Fig.~\ref{single_comm} illustrates the received signal-to-noise ratio (SNR) at the user versus its azimuth angle, with the zenith angle fixed at $\theta=0^\circ$. As the user's azimuth angle varies from $-\pi/3$ to $\pi/3$, the RA dynamically adjusts its boresight direction to track and align with the user's direction, thus maintaining a stable received SNR. In contrast, for the fixed-antenna-based transmitter, user movement results in a growing misalignment between the antenna boresight and the user direction, thus leading to a significant degradation in the received SNR. The above experimental results validate the effectiveness of RA for enhancing communication coverage performance.

\subsection{Simulation Results: Multi-User Communication and Multi-Target Sensing}
\begin{figure}[!t]
	\centering
	\includegraphics[width=3.3in]{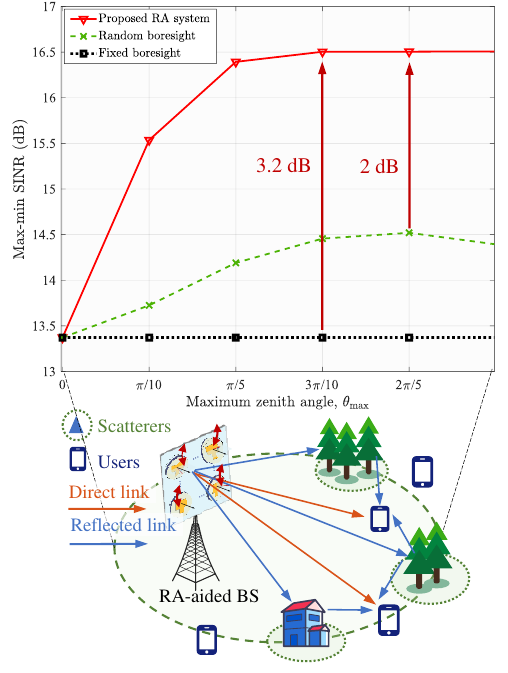}
	\vspace{-0.3cm}
	\caption{Max-min SINR of different schemes versus the maximum zenith angle of RA.}
	\label{multi_comm}
\end{figure}
\begin{figure}[!t]
	\centering
	\includegraphics[width=3.3in]{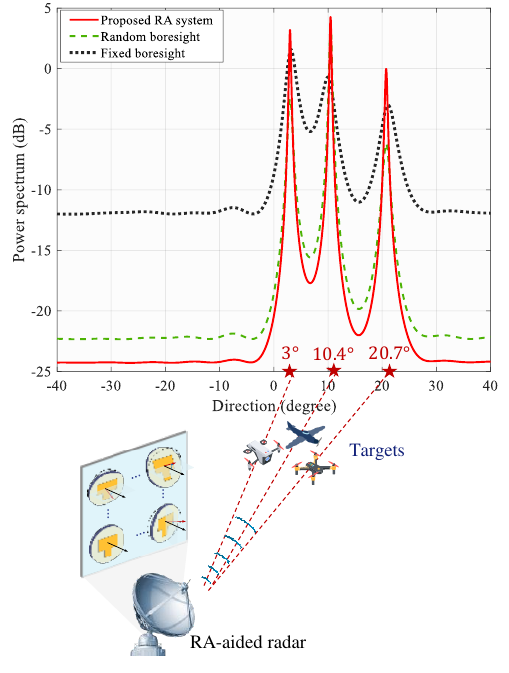}
	\caption{Spatial power spectrum comparison of different schemes.}
	\label{multi_sensing}
\end{figure}
	Next, we consider a multi-user communication system (with four users under multi-path channels) and a multi-target sensing system (involving three targets under LoS channels) operating at a carrier frequency of $2.4$ GHz. The BS and radar are equipped with uniform planar arrays (UPAs) comprising $N=4\times4$ directional RAs with the antenna radiation pattern given in \cite{zheng2025rotatable}. The users/targets are randomly distributed within a semicircular region centered at the BS/radar with a radius of 50 meters. Furthermore, to highlight the advantages of the optimized RA systems, we compare them against two baseline systems: 1) Random boresight, where each antenna's boresight direction is randomly generated within its feasible boresight range; and 2) Fixed boresight, where each antenna's boresight is aligned with its default (normal) direction.

    Fig.~\ref{multi_comm} shows the max-min signal-to-interference-plus-noise ratio (SINR) versus the maximum zenith angle $\theta_{\mathrm{max}}$ of RA in the multi-user communication system. As $\theta_{\mathrm{max}}$ increases, the RA system gains more spatial flexibility to adapt its radiation pattern to the spatial distribution of users, thus boosting the received SINR more significantly. Moreover, the proposed RA system consistently outperforms both baseline schemes across all values of $\theta_{\mathrm{max}}$, achieving up to $2$ dB and $3.2$ dB improvements compared to random boresight and fixed boresight, respectively. Notably, a significant SINR improvement is observed even when $\theta_{\mathrm{max}}\le\pi/10$, highlighting that substantial performance gains can be achieved with only a small or moderate range of boresight adjustment when properly optimized.

    Fig.~\ref{multi_sensing} shows the sensing performance of different schemes for multi-target sensing in terms of spatial power spectrum. It is observed that compared to the random and fixed boresight schemes, the proposed RA-enabled sensing achieves much sharper peaks in the power spectrum aligned with the target directions while maintaining significantly lower sidelobe levels. This demonstrates the enhanced angular resolution and sensing accuracy attributed to the increased directional gain provided by the optimized RA boresight directions.

\section{Conclusions}
By exploiting the appealing 3D boresight adjustment of antennas, RA offers additional DoFs to cost-effectively enhance wireless communication and sensing performance without requiring additional antenna resources or extra deployment space. This article has provided an overview of RA including its applications, fundamentals, challenges, and promising solutions as well as its significant advantages over conventional fixed-antenna architectures. Case studies have also been presented to demonstrate the substantial performance gains achieved by RA-enabled wireless systems.
Looking ahead, we envision that future RAs will become increasingly ``smart'', capable of automatically scanning, sensing, tracking, and learning from wireless environments\textemdash much like the function of human eyes.
This evolution will empower advanced directional and flexible communications for simultaneously enhancing signal strength and mitigating interference, thus paving the way for more agile, efficient, and intelligent wireless networks.

\section*{Acknowledgment}
The work of Beixiong Zheng was supported in part by the National Natural Science Foundation of China under Grant 62571193, Grant 62201214, and Grant 62331022, the Natural Science Foundation of Guangdong Province under Grant 2023A1515011753, the Guangdong program of under Grant 2023QN10X446 and Grant 2023ZT10X148, the GJYC program of Guangzhou under Grant 2024D01J0079 and Grant 2024D03J0006, and the Fundamental Research Funds for the Central Universities under Grant 2024ZYGXZR087.
This work of Changsheng You was supported in part by the National Natural Science Foundation of China under Grant 62571227; in part by the Shenzhen Science and Technology Program under Grant JCYJ20240813094212016; and in part by the Program under Grant 2023QN10X152.
	
\bibliographystyle{IEEEtran}
\bibliography{ref}

\section*{Biographies}
\noindent{{\bf Beixiong Zheng}
	[SM] (bxzheng@scut.edu.cn) is an Associate Professor with the School of Microelectronics, South China University of Technology, Guangzhou, China.
}
\\

\noindent{{\bf Tiantian Ma}
	(mitiantianma@mail.scut.edu.cn) is with the School of Microelectronics, South China University of Technology, Guangzhou, China.
}
\\

\noindent{{\bf Changsheng You}
	 (youcs@sustech.edu.cn) is an Assistant Professor with the Department of Electronic and Electrical Engineering, Southern University of Science and Technology,
	China.
}
\\

\noindent{{\bf Jie Tang}
	[SM] (eejtang@scut.edu.cn) is a Professor with the School of Electronic and Information Engineering, South China University of Technology, Guangzhou, China.
}
\\

\noindent{{\bf Robert Schober}
	[F] (robert.schober@fau.de) is an Alexander von Humboldt Professor and the Chair for Digital Communication at Friedrich-Alexander University of Erlangen-Nuremberg (FAU), Germany.
}
\\

\noindent{{\bf Rui Zhang}
	[F] (elezhang@nus.edu.sg) is a Provost's Chair  Professor with the Department of Electrical and Computer Engineering of National University of Singapore, Singapore. 
}
	
\end{document}